\documentclass{jfm}

\usepackage{graphicx}
\usepackage{natbib}

\usepackage{amssymb}
\usepackage{amsmath}
\usepackage{subfig}

\newcommand{\mr}[1]{\mathrm{#1}}
\newcommand{\pd}{\partial}
\newcommand{\dd}{\mathrm{d}}
\newcommand{\ii}{\mathrm{i}}

\newcommand{\und}{\quad \text{and}\quad}
\newcommand{\for}{\quad\text{for}\quad}
\newcommand{\reel}{\mathrm{Re}}
\newcommand{\imag}{\mathrm{Im}}

\ifCUPmtlplainloaded \else
  \checkfont{eurm10}
  \iffontfound
    \IfFileExists{upmath.sty}
      {\typeout{^^JFound AMS Euler Roman fonts on the system,
                   using the 'upmath' package.^^J}%
       \usepackage{upmath}}
      {\typeout{^^JFound AMS Euler Roman fonts on the system, but you
                   dont seem to have the}%
       \typeout{'upmath' package installed. JFM.cls can take advantage
                 of these fonts,^^Jif you use 'upmath' package.^^J}%
      }
  \else
  \fi
\fi

\ifCUPmtlplainloaded \else
  \checkfont{msam10}
  \iffontfound
    \IfFileExists{amssymb.sty}
      {\typeout{^^JFound AMS Symbol fonts on the system, using the
                'amssymb' package.^^J}%
       \usepackage{amssymb}%

      }{}
  \fi
\fi

\ifCUPmtlplainloaded \else
  \IfFileExists{amsbsy.sty}
    {\typeout{^^JFound the 'amsbsy' package on the system, using it.^^J}%
     \usepackage{amsbsy}}
    {\providecommand\boldsymbol[1]{\mbox{\boldmath $##1$}}}
\fi

\newcommand\Rey{\mbox{\textit{Re}}}  

\newsavebox{\astrutbox}
\sbox{\astrutbox}{\rule[-5pt]{0pt}{20pt}}

\title[Flow over a cavity containing a second immiscible fluid.]{Longitudinal and transversal flow over a cavity containing a second immiscible fluid}

\author[C. Sch\"onecker %
and S. Hardt]%
{C\ls L\ls A\ls R\ls I\ls S\ls S\ls A\ns S\ls C\ls H\ls \"O\ls N\ls E\ls C\ls K\ls E\ls R%
  \thanks{Email address for correspondence: schoenecker@csi.tu-darmstadt.de},\ns
\and S\ls T\ls E\ls F\ls F\ls E\ls N\ns H\ls A\ls R\ls D\ls T}

\affiliation{Institute for Nano- and Microfluidics, Center of Smart Interfaces, Technische Universit\"at Darmstadt, 64287 Darmstadt, Germany\\[\affilskip]
}

\pubyear{2010}
\volume{650}
\pagerange{119--126}
\date{?; revised ?; accepted ?. - To be entered by editorial office}

\begin{document}

\maketitle

\begin{abstract}
An analytical solution for the flow field of a shear flow over a rectangular cavity containing a second  immiscible fluid is derived. While flow of a single-phase fluid over a cavity is a standard case investigated in fluid dynamics, flow over a cavity which is filled with a second immiscible fluid, has received little attention. The flow filed inside the cavity is considered to define a boundary condition for the outer flow which takes the form of a Navier slip condition with locally varying slip length. The slip-length function is determined from the related problem of lid-driven cavity flow. Based on the Stokes equations and complex analysis it is then possible to derive a closed analytical expression for the flow field over the cavity for both the transversal and the longitudinal case. The result is a comparatively simple function, which displays the dependence of the flow field on the cavity geometry and the medium filling the cavity. The analytically computed flow field agrees well with results obtained from a numerical solution of the Navier-Stokes equations. The studies presented in this article are of considerable practical relevance, for example for the flow over superhydrophobic surfaces. 
\end{abstract}


\section{Introduction}
Flows in and over cavities are important blueprints for many phenomena in fluid dynamics. On a reduced scale, they illustrate phenomena occurring in various, often more complex applications. These might for example be in aerodynamics, where the vortices forming inside cavities or grooves in surfaces are studied \citep{pan1967,moffat1963, higdon1985, shankar2000}. The corresponding flow patterns affect the drag a fluid streaming over the surface is subjected to \citep{shen1985, prasad1989}. Traditionally, such troughs, or more generally surface roughness features, have always been perceived as unwanted, since the vortices forming inside the cavities may lead to higher drag or acoustic disturbances. Yet, at higher Reynolds numbers, riblets may also reduce drag, which has been of great interest more recently \citep{viswanath2002, jovanovic2010}.

In all these investigations, only a single fluid phase flowing over the cavity and filling it at the same time is considered. However, there exist numerous applications, where a fluid flows over a cavity filled with a second immiscible fluid. It is well known that by capillary condensation vapor condenses in sufficiently small capillaries or channels although the vapor pressure is below the bulk saturation pressure \citep{fisher1979, fisher1981, cicciotti2008}. So does vapor inside the roughness features of a surface. This may have an immediate impact on the applications in aerodynamics mentioned above, since the condensed water has a viscosity different from the gas and hence moves at a slower speed. Moreover, we can also encounter the inverse situation, where small bubbles, often of nanometer size, are enclosed in the roughness features of the wall of a water-filled channel \citep{yang2007}. Additionally, all superhydrophobic, Lotus leaf-mimicking surfaces take advantage of such enclosed air \citep{lee2008, ma2006, nosonovsky2008, rothstein2010}. Other examples are the flow over membranes or porous media, which may be filled with a second  immiscible fluid as well.

Besides the work on flows over cavities, which includes the fluid motion inside the cavity, there is also investigation of flows over surfaces with periodically patterned boundary conditions. In such a context a cavity is usually represented by a certain boundary condition to the flow above it. The flow inside is not considered. \citet{philip1972} analyzes the flow over a no-shear slot within a no-slip wall in longitudinal and transverse direction. A no-shear interface is certainly the ideal limit, which cannot be reached by a real viscous fluid. Although this condition is often used for modeling air-liquid interfaces \citep{lauga2003, sbragaglia2007, Davis2009, crowdy2010, davies2010}, it is not immediately clear how well air fulfills these ideal characteristics.

The no-shear condition corresponds to an infinite slip length $\beta$ in the Navier boundary condition \citep{navier1823}
\begin{equation}
\left.u\right|_\mr{boundary}=\beta\left.\frac{\pd u}{\pd y}\right|_\mr{boundary},
\label{eq:navierbc}
\end{equation}
which allows for a nonzero velocity $u$ at the boundary. The slip length $\beta$ relates the velocity to its gradient in the direction normal to the surface $\frac{\pd u}{\pd y}$ and can be understood as an imaginary depth below the surface at which the velocity extrapolates to zero. Still, when employing equation \eqref{eq:navierbc} for the modeling of a fluid-filled cavity, it is by no means clear how to specifically choose $\beta$. \citet{belyaev2010} have calculated a flow over an array of slots with constant slip length. However, it remains to unclear how $ \beta $ is related to the slot geometry or to the properties of the fluid filling the cavities. Furthermore, in reality the slip length will not be constant but a spatially varying quantity. So far, \citet{hocking1975} has briefly addressed the effect of the viscosity of the medium inside a cavity, but only found an approximation to the streamlines in terms of an infinite Fourier series in the case of an infinite slip length.

In this paper, a representation of a rectangular cavity in the form of a boundary condition for a second immiscible fluid flowing over a surface is derived. It is directly related to the flow pattern inside the cavity, i.e. to the geometry of the cavity and the viscosity of the fluid it contains. The effect of these properties on the outer flow can therefore be directly explored.

Based on this, we calculate the flow field of a shear flow directed transversely and longitudinally with respect to the cavity. The solution to this mixed boundary value problem can be achieved though a representation of the flow by complex variables. Unlike with often employed methods based on Fourier series, we can derive a closed-form expression and present a simple, explicit formula for the flow field.

\section{Model of the cavity and boundary conditions} \label{modeling}
\begin{figure}
  \centerline{\includegraphics[width=0.7\textwidth]{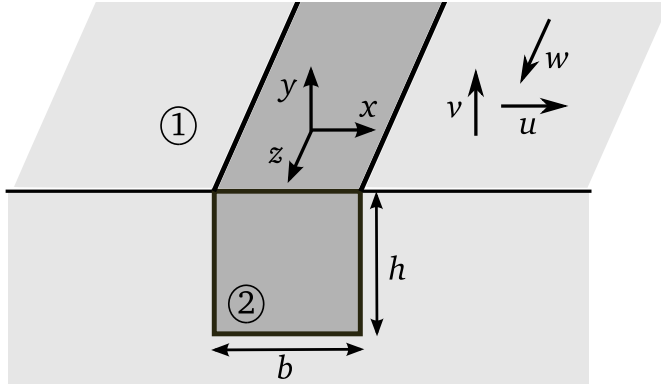}}
  \caption{Schematic of the considered cavity with coordinate and velocity directions}
\label{fig:kavitaet}
\end{figure}

We consider a rectangular cavity of depth $h$, width $b$ and infinite extension in $z$-direction as depicted in fig. \ref{fig:kavitaet}. Fluid 1 flows over the cavity, while fluid 2 fills the same. The interface between both fluids is assumed to be flat. For small cavities, this is usually justified, since surface tension provokes a flattening of the interface and will only permit negligibly small curvatures \citep{higdon1985}. In more general terms it means that we consider a flow problem at sufficiently small capillary numbers.

In the following, a boundary condition is derived, describing the effect of the cavity on the outer flow. It can be employed without the need of calculating the detailed flow inside the cavity.

A first-principles modeling of the cavity is achieved by taking into account the viscosity of the medium inside the cavity and the geometry of the cavity. The modeling is exemplified for a flow perpendicular to the cavity. The longitudinal case is analogous. 

At the fluid-fluid interface, continuity of velocity and shear stress applies:
\begin{equation}
\left.
\begin{aligned}
u_{1}&=u_2\\
\eta_1\frac{\pd u_{1}}{\pd y}&=\eta_2\frac{\pd u_{2}}{\pd y}
\end{aligned}
\right\} \for |x|<\frac{b}{2},\quad y=0,
\label{eq:transitioncond}
\end{equation}
while at the wall, the no-slip condition holds:
\begin{equation}
u_1 =0 \for |x|>\frac{b}{2},\quad y=0.
\end{equation}
$u$ denotes the velocity in $x$-direction %
and $\eta$ is the dynamic viscosity. The indices 1 and 2 refer to the respective fluids. %

For the velocity $u_2$ at the upper side of the cavity a Navier slip condition %
\begin{equation}
u_2(x,0)=\delta(x)\left.\frac{\pd u_{2}}{\pd y}\right|_{y=0}
\end{equation}
holds, with the local slip length function $\delta(x)$ to be determined. Incorporating this equation, the two transition conditions \eqref{eq:transitioncond} collapse into one boundary condition, such that the boundary conditions for fluid 1 are
\begin{align}
u_{1}&=N \delta(x)\frac{\pd u_{1}}{\pd y} &\for |x|<\frac{b}{2},\quad y&=0 \label{eq:bcreel1}\\
u_1&=0 &\for |x|>\frac{b}{2},\quad y&=0  \label{eq:bcreel2}
\end{align}
with $N$ abbreviating the viscosity ratio of the two fluids
\begin{equation}
N=\frac{\eta_1}{\eta_2}.
\end{equation}
Equation \eqref{eq:bcreel1} again has the form of a Navier slip condition. The net slip length for fluid 1 is thus the product of the viscosity ratio and the slip length  function.

As soon as the slip length function has been determined, the flow over the cavity can now be studied without the need of computing the flow field inside the cavity. This is of great advantage for analytical calculations. Since in the following the focus is on the flow over the surface outside of the cavity, the subscript 1 is no longer needed and is dropped from now on.

A suitable expression for the slip-length function $ \delta (x) $ needs to be found.  At the corners $x=\pm\frac{b}{2}$, obviously $\delta(x)=0$ to make both boundary conditions match. At the center of the cavity at $x=0$, the flow is least suppressed by the walls of the cavity, and hence, the slip length reaches a maximum $d$. 

Values for $d$ can be obtained from the investigation of lid driven cavities. Here, the upper boundary of the cavity is also considered being flat. \citet{pan1967} and \citet{shankar2000} have shown that, depending on the aspect ratio of the cavity, one or more counter-rotating vortices of decreasing intensity develop in the cavity. For low Reynolds numbers, the centers of these vortices are always located on the centerline of the cavity $x=0$. The distance from the center of the first of these vortices to the top of the cavity $d^*$ gives an upper bound to $d$. This is due to the shape of the velocity field (cf. \citet{higdon1985}) being driven by the upper boundary. \citet{pan1967} give the center of the first vortex $D=d^*/b$ as a function of the aspect ratio $h/b$ (see table \ref{tab:d}), which rapidly approaches one fourth of the cavity width as the aspect ratio increases.

\begin{table}
  \begin{center}
\def~{\hphantom{0}}
  \begin{tabular}{lllllll}
      $h/b$  &  &0.25   &  0.50  & 1.0 & 2.0 & 5.0 \\
			$d^*/b$ & &0.0875  & 0.1628 & 0.24 & 0.25 &0.25\\
  \end{tabular}
  \caption{Center of the first vortex after \citet{pan1967}}
  \label{tab:d}
  \end{center}
\end{table}

This rather heuristic approach for determining $ \delta(x) $ can be formalized by introducing iterative maps
\begin{equation}
\Gamma_1 : U_i (x) \rightarrow \delta_i (x)
\end{equation}
and 
\begin{equation}
\Gamma_2 : \delta_i (x) \rightarrow U_{i+1} (x).
\end{equation}
$ \Gamma_1 $ denotes that from the solution of the generalized lid-driven cavity problem with a velocity distribution $ U_i (x) $ prescribed at the upper boundary of the cavity a slip-length function $ \delta_i (x) $ is determined. $ \Gamma_2 $ maps a slip-length function used as a boundary condition at the top of the cavity to the velocity distribution at this boundary. Successive application of the map $ \Gamma_1 \circ \Gamma_2 $ generates a series of approximations $ \delta_i (x), (i = 1, 2, \dotsc) $ to the slip-length function $ \delta(x) $, i.e. $  \lim_{i \to \infty} \delta_i (x) = \delta(x) $, provided that the iteration scheme converges. Below we will show that when starting with $ U_1 (x) = \mr{const.} $, already $ \delta_1 (x) $ -- determined in a simplified manner -- yields a good approximation to the flow field above the cavity. Therefore, in the framework of this article no further iterations will be performed.    

In that spirit, we utilize the results of \citet{pan1967} to choose a differentiable function $\delta(x)$ with $\delta(\pm\frac{b}{2})=0$ and $\delta(0)=d^*$. The specific form of $\delta(x)$ is still to be determined. This will be done in a way that allows calculating a closed-form solution for the flow field, as will become clear in the following section.

\section{Transverse Flow}
In the following, a fluid flow in $x$-direction over the cavity is investigated and its flow field is derived. The fluid is driven by a shear stress $\tau_\infty$ in $x$-direction at $ y \rightarrow \infty $.

At sufficiently small Reynolds numbers the flow obeys the Stokes equations
\begin{align}
 -\frac{\pd p}{\pd x}+\frac{\pd^2 u}{\pd x^2}+\frac{\pd^2 u}{\pd y^2}=&0\\
 -\frac{\pd p}{\pd y}+\frac{\pd^2 v}{\pd x^2}+\frac{\pd^2 v}{\pd y^2}=&0.
\end{align}
$u$ and $v$ are the velocities in $x$- and $y$-direction, respectively, and $p$ is the pressure. With the stream function
\begin{equation}
\frac{\pd \Psi}{\pd x}=-v \quad \text{and}\quad \frac{\pd \Psi}{\pd y}=u
\end{equation}
the Stokes equations transform into the biharmonic equation
\begin{equation}
\Delta \Delta \Psi=0.
\end{equation}

The solution to the biharmonic equation is given by Goursat's theorem \citep{smirnow_b_32, muskhelishvili_b_elasticity}, a special form of which reads
\begin{equation}
\Psi=\reel\left((\bar{\mathfrak{z}}-\mathfrak{z})W_\mr{t}(\mathfrak{z})\right).
\label{eq:goursat}
\end{equation}
The Goursat function $W_\mr{t}(\mathfrak{z})$ is an analytic function to be found, with the subscript t indicating the transverse flow direction. $\reel$ is the real part and the complex coordinate is given by
\begin{equation}
\mathfrak{z}=x+ \ii y \und \bar{\mathfrak{z}}=x- \ii y.
\end{equation}
In equation \eqref{eq:goursat}, the general Goursat theorem containing two Goursat functions has been reduced to an expression containing only one unknown function by elimination of the undetermined constant in the stream function. Due to the impermeability of walls lying on the real axis, the stream function is set to 0 at these boundaries \citep[cf.][]{philip1972,garabedian1966}. Expression \eqref{eq:goursat} has previously been used in the solution of similar problems \citep{atanackovic1977, philip1972}.

With the ansatz \eqref{eq:goursat} the boundary conditions \eqref{eq:bcreel1} and \eqref{eq:bcreel2} read
\begin{align}
\imag(W_\mr{t})&=2 N \delta(x)\reel(W'_\mr{t}) &\for |x|&<\frac{b}{2},\quad y=0  \label{eq:bcimag1}\\
\imag(W_\mr{t})&=0 &\for |x|&>\frac{b}{2},\quad y=0.  \label{eq:bcimag2}
\end{align}
The prime denotes the derivative with respect to $\mathfrak{z}$.

\begin{figure}
\centering
	\includegraphics[width=0.7\textwidth]{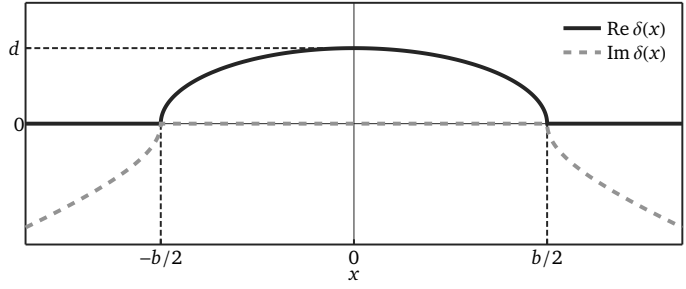}
	\caption{real (solid line) and imaginary (dashed line) part of $\delta(x)$}
	\label{fig:slip}
\end{figure}

As an elementary mathematical case, the slip length function $\delta(x)$ is assumed to be of elliptic form, fulfilling the conditions of section \ref{modeling}. In complex notation $\delta(x)$ can be written as
\begin{equation}
\delta(x)=-\ii 2 D \sqrt{x^2-\frac{b^2}{4}}
\label{eq:elliptslip}
\end{equation}
being real for $|x|<b/2$ and imaginary for $|x|>b/2$ (fig. \ref{fig:slip}). This form is essential for keeping the following calculations analytically tractable. Nevertheless, one is not restricted to an ellipse. \citet[p.68-69]{cherepanov_b_brittle} outlined a method of approximating any arbitrary function being continuous on a finite interval and vanishing at its ends according to the same requirements concerning its real and imaginary parts as given above. However, for convenievce we will use the expression of eq. \eqref{eq:elliptslip}.

Similarly to the boundary conditions at $y=0$, the condition of a given shear stress $\tau_\infty=\eta_1\,\pd u/\pd y$ at infinity can be written in term of $W(\mathfrak{z})$ as
\begin{equation}
\reel(W'_\mr{t})= \frac{\tau_\infty}{4\eta_1} \for y \rightarrow \infty.
\label{eq:bcimaginf}
\end{equation}

The solution procedure follows the method of \citet{sotkilava1974}, originally employed in elasticity theory. An auxiliary function
\begin{equation}
F_\mr{t}(\mathfrak{z})=W'_\mr{t}(\mathfrak{z})-\frac{1}{4 D N \sqrt{\mathfrak{z}^2-\frac{b^2}{4}}} W_\mr{t}(\mathfrak{z})
\label{eq:F}
\end{equation}
is chosen such that the boundary conditions \eqref{eq:bcimag1} and \eqref{eq:bcimag2} take the form
\begin{align}
\reel(F_\mr{t}(\mathfrak{z}))&=0 &\for |x|<\frac{b}{2},\quad y&=0 \label{eq:bcf1}\\
\imag(F_\mr{t}(\mathfrak{z}))&=0 &\for |x|>\frac{b}{2},\quad y&=0. \label{eq:bcf2}
\end{align}
The condition at infinity \eqref{eq:bcimaginf} is
\begin{equation}
\reel(F_\mr{t}(\mathfrak{z}))=\frac{\tau_\infty}{4\,\eta_1}\left(1-\frac{1}{4 D N}\right) \for y \rightarrow \infty.
\label{eq:bcinff}
\end{equation}
The conditions \eqref{eq:bcf1} and \eqref{eq:bcf2} constitute a mixed boundary value problem for the half plane. Its solution is given by the Keldysh-Sedov formula \citep{keldysh1937, muskhelishvili_b_integral}. Since  \eqref{eq:bcf1} and \eqref{eq:bcf2} are homogeneous, only the homogeneous part of the Keldysh-Sedov formula is required, which reads
\begin{equation}
F_\mr{t}(\mathfrak{z})=\frac{L(\mathfrak{z})}{\sqrt{\mathfrak{z}^2-\frac{b^2}{4}}}.
\label{eq:keldyshsedov}
\end{equation}
In our case, $L(\mathfrak{z})$ is a linear function with two real coefficients \citep{lawrentjew1967} to be determined. This is achieved by requiring the condition at infinity \eqref{eq:bcinff}, giving
\begin{equation}
F_\mr{t}(\mathfrak{z})=\frac{\tau_\infty}{4 \eta_1}(1-\frac{1}{4 D N})\frac{\mathfrak{z}}{\sqrt{\mathfrak{z}^2-\frac{b^2}{4}}}.
\end{equation}

The auxiliary differential equation \eqref{eq:F} can now be solved. Its general solution is
\begin{equation}
W_\mr{t}(\mathfrak{z})=W_0(\mathfrak{z})\left(C+\int\limits_{\mathfrak{z}_0}^\mathfrak{z} F_\mr{t}(\mathfrak{Z}) W_0^{-1}(\mathfrak{Z}) \dd \mathfrak{Z}\right)
\label{eq:dgllsg}
\end{equation}
with the solution to the homogeneous problem
\begin{equation}
W_0(\mathfrak{z})=e^{\int\limits_{\mathfrak{z}_0}^\mathfrak{z} \frac{1}{4 D N  \sqrt{\mathfrak{Z}^2-\frac{b^2}{4}}}\dd \mathfrak{Z}}
\label{eq:w0allg}
\end{equation}
and a constant $C$ to be determined.
$C$ can be obtained from the condition of symmetry
\begin{equation}
\imag(W'_\mr{t}(\mathfrak z))=0 \quad \text{at}\quad \mathfrak{z}=0,
\label{eq:smmetriebed}
\end{equation}
ensuring the analycity of $W_\mr{t}(\mathfrak{z})$. 

Integrating from $\mathfrak{z}_0=\frac{b}{2}$, i.e. evaluating the Cauchy principal value of the integrals in \eqref{eq:dgllsg} and \eqref{eq:w0allg}, the result for $W_\mr{t}(\mathfrak{z})$ is computed as
\begin{equation}
W_\mr{t}(\mathfrak{z})=\frac{\tau_\infty}{4 \eta_1} \frac{ \mathfrak{z}+ 4 D N \sqrt{\mathfrak{z}^2-\frac{b^2}{4}}}{1+4 D N}.
\end{equation}
Following Goursat's theorem \eqref{eq:goursat}, the expression for the streamfunction is
\begin{equation}
\Psi=\frac{\tau_\infty}{2 \eta_1} \frac{y}{1+4 D N}\left(y+4 D N \imag{\sqrt{(x+\ii y)^2-\frac{b^2}{4}}}\right).
\label{eq:stromf_transv}
\end{equation}
This result is consistent with the work of \citet{philip1972}, who calculated the transverse flow over a no-shear slot. In the limit of a very low viscosity fluid in the cavity
\begin{equation}
\lim_{N\rightarrow\infty}\Psi=\frac{\tau_\infty}{2 \eta_1}\, y \,\imag{\sqrt{(x+\ii y)^2-\frac{b^2}{4}}},
\end{equation}
which is equivalent to Philip's result.

When evaluating \eqref{eq:stromf_transv} computationally in the second quadrant of $z$ one should be aware of crossing a branch cut, which for a square root is usually chosen along the negative real axis. To avoid ambiguity one may want to write $\ii \sqrt{\frac{b^2}{4}-\mathfrak{z}^2}$ instead of $\sqrt{\mathfrak{z}^2-\frac{b^2}{4}}$ for $x<0$.

\begin{figure}
\centering
	\includegraphics{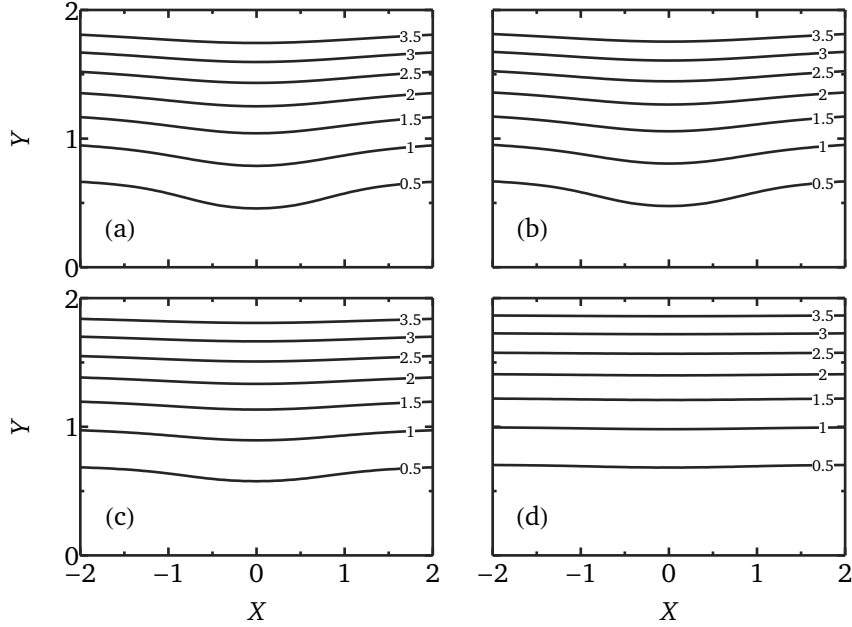}
	\caption{Transverse flow streamline patterns $\tilde \Psi$ for $D=0.25$ and varying viscosity ratio: a) $N=100$, b) $N=10$, c) $N=1$, d) $N=0.1$}
	\label{fig:stroml_trans}
\end{figure}

In fig. \ref{fig:stroml_trans}, the streamlines are plotted in the form
\begin{equation}
\tilde \Psi=\Psi \frac{2 \eta_1}{\tau_\infty}= \frac{Y}{1+4 D N}\left(Y+4 D N \imag{\sqrt{(X+\ii Y)^2-1}}\right)
\label{eq:stromf_transv_ed}
\end{equation}
with $X= x \frac{2}{b}$ and $Y= y \frac{2}{b}$. The only flow-determining parameters are $D$, the ratio of the maximum local slip length and the cavity width, and the viscosity ratio $N$. Fig. \ref{fig:stroml_trans} shows the streamlines for cavities deeper than wide ($D=0.25$) and varying $N$. Clearly, the higher the viscosity ratio, i.e. the lower the viscosity of the cavity medium, the more the presence of the cavity influences the flow over the wall. For a high viscosity ratio, the flow is notably accelerated towards the cavity, while for a small viscosity ratio the flow pattern approaches the one of a plain wall.

\begin{figure}
\centering
	\subfloat[][]{%
	\includegraphics[width=0.48\textwidth]{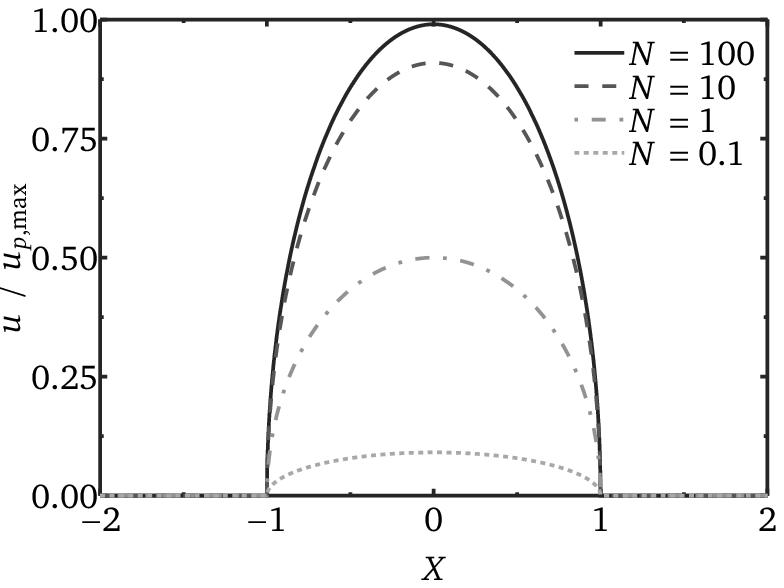}}
	\hspace{5pt}%
\subfloat[][]{%
	\includegraphics[width=0.48\textwidth]{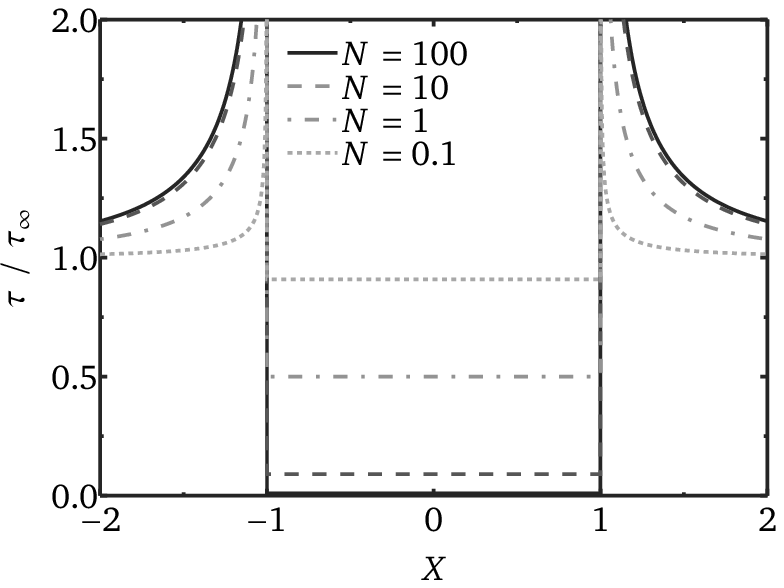}}
	\caption{Normalized a) velocity and b) shear stress along the boundary $y=0$}
	\label{fig:rbtrans}
\end{figure}

This behavior is explicitly apparent from fig. \ref{fig:rbtrans}. The velocity $u$ at the boundary $y=0$ is normalized with $u_\mr{p,max}$, being the maximum achievable velocity for the perfect slip case $N\rightarrow\infty$ at $x=y=0$. The velocity obviously increases with $N$. In the same manner, the shear stress at $y=0$ is similar to the average shear stress for low $N$, while being reduced towards 0 for high $N$. It can be further observed, that for the chosen elliptic slip length distribution $\delta(x)$, the shear stress is constant along the cavity.

It should be noted that for $N=1$ the flow field is not identical to the case of a single-phase flow. The flow profile is modified due to the flat fluid-fluid interface, which strictly separates both fluids and cannot be crossed.

In the case of water flowing over an air-containing cavity, the maximum velocity at $x=0$ is $u_0/u_\mr{p,max}=0.9823$ and the shear stress is $\tau/\tau_\infty=0.0177$ for $D=0.25$. These numbers illustrate the accuracy of previous models, where a vanishing shear stress is assumed at the air-water interface. For cavities deeper than wide, the maximum velocity was hence calculated about 2\% too high, which nevertheless, is still a reasonable approximation. However, for smaller aspect ratios of the cavity, the assumption of vanishing shear stress becomes increasingly inappropriate, e.g. for $h/b=0.25$, i.e. $D=0.0875$ from \citet{pan1967}, the deviation is about 5\%.

\begin{figure}%
\centering
\includegraphics[width=0.5\textwidth]{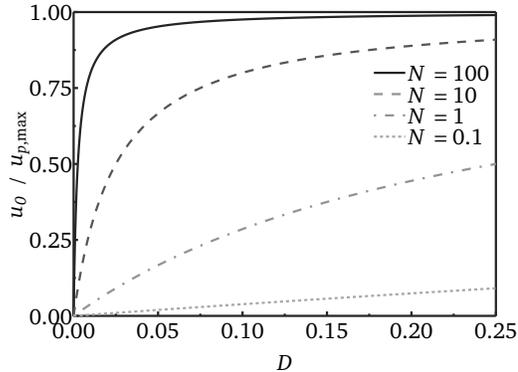}%
\caption{Variation of the velocity at $\mathfrak z=0$ with the slip parameter $D$ for a range of viscosity ratios $N$}%
\label{fig:trans_klTiefen}%
\end{figure}

For lower viscosity ratios than for the water-air system ($N\approx 56$), the influence of the cavity depth gets more prominent, which can be observed from fig. \ref{fig:trans_klTiefen}. Here, the dependence of the normalized velocity $u_0/u_\mr{p,max}$ at $\mathfrak z=0$ on the slip parameter $D$, which is a function of the cavity height, is shown for various values of $N$. At $N=1$, $u_0/u_\mr{p,max}$ drops down to about 50\% when the aspect ratio is lowered from $h/b>1$ to $h/b=0.25$. In the case of air flowing over water ($N=0.018$) $u_0/u_\mr{p,max}$ is reduced to approx. one third upon the same variation of the aspect ratio. Furthermore, it can be observed, that the influence of a variation of $N$ on $u_0/u_\mr{p,max}$, translating to a modification of the outer flow field, is strongest around $N=1$. At very large or small values of $N$, a variation of $N$ has only little effect. This fact is also apparent in fig. \ref{fig:stroml_trans} (a) and (b), which differ only slightly. 

As explained above, $\delta(x)$ may be a general function with $\delta(\pm\frac{b}{2})=0$ and a maximum $\delta(0)=d$. As a minimum model and in order to keep the calculations analytically tractable, we have employed an elliptic form with $d$ taking the value of its upper bound known from lid driven cavity flows. In order to assess the accuracy of this model, its results are compared to numerical calculations. 

\begin{figure}
\centering
	\subfloat[][]{%
	\includegraphics[width=0.49\textwidth]{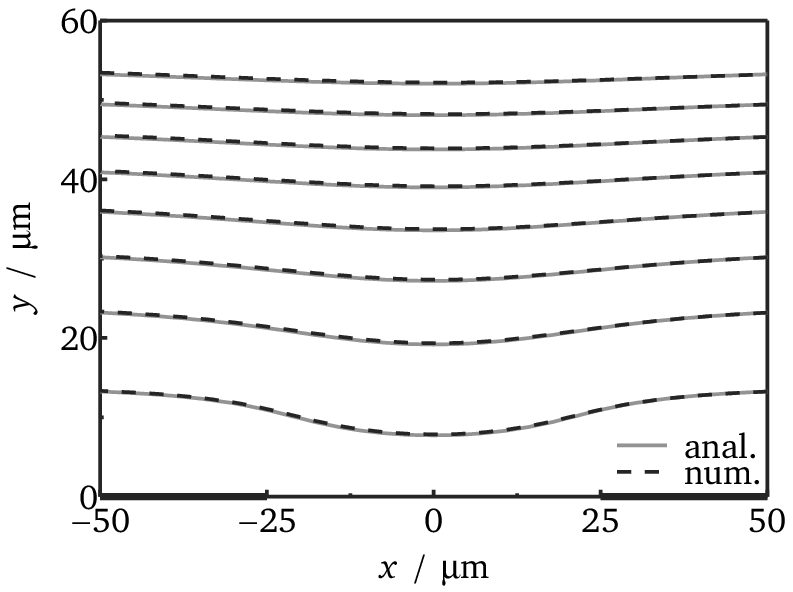}}
	\hspace{5pt}%
\subfloat[][]{%
	\includegraphics[width=0.49\textwidth]{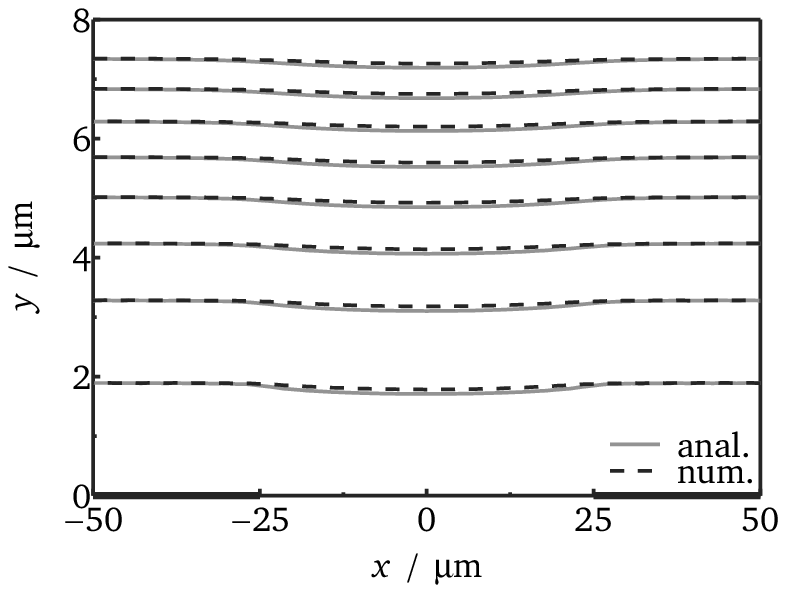}}
	\caption{Comparison of the transverse flow streamlines calculated analytically with equation \eqref{eq:stromf_transv} and numerically: a) water flowing over an air-filled cavity b) air flowing over a water-filled cavity. Note the different scalings of the $y$-axis in parts (a) and (b)}
	\label{fig:comsolvgl_trans2}
\end{figure}

Both the fluid inside the cavity as well as the fluid flowing over the cavity are included in the numerical computation. Exemplary, water (density $\rho=10^3 \,\mr{kg/m^3}$ and dynamic viscosity $\eta=10^{-3} \,\mr{Pa \, s}$) and air ($\rho=1.2 \,\mr{kg/m^3}$, $\eta=1.8 \cdot 10^{-5} \,\mr{Pa \, s}$) have been considered. At the interface between the fluids, continuity of velocity and stress \eqref{eq:transitioncond} holds. The dimensions of the cavity are $b=h=50\,\umu\mr m$. At a distance to the wall of $y=500\,\umu\mr m$, a shear stress of $0.1\,\mr{Pa}$ is imposed, so that $\Rey=\left.u\right|_{x=0,y=0} b \rho/\eta\ll1$ for both fluids. The Navier-Stokes equations obeying these conditions have been solved with  the commercial finite-element software 
COMSOL Multiphysics (COMSOL Multiphysics GmbH, G\"ottingen, D) on a calculation domain of 3\,mm width with a triangular mesh of approx. 65000 elements. The mesh was strongly refined along the fluid-fluid interface to accomodate the discontinuous change in the boundary condition for the flow over the cavity between the fluid-solid and fluid-fluid interfaces. Subsequently, the numerically computed velocity was interpolated and integrated to yield the stream function. In fig. \ref{fig:comsolvgl_trans2}, a comparison between analytical and numerical results is displayed for two different cases: water flowing over an air-filled cavity (a) and air flowing over a water-filled cavity (b). In case (a), an excellent agreement between the analytical and the numerical results is found. The agreement is still good in case (b), only upon great magnification of a region close to the air-water interface some deviations between the analytical and the numerical results become visible, c.f. fig \ref{fig:comsolvgl_trans2} (b). Specifically, the analytically computed streamlines have a higher curvature than the numerically computed ones. This corresponds to a small overprediction of the velocity, which is expected owing to the choice of $d$ being an upper bound. 

\section{Longitudinal Flow}
We now address the longitudinal shear flow over a cavity. The problem is mathematically closely related to the transverse flow problem. For an infinitely long cavity, the velocity $w$ in $z$-direction is governed by the Laplace equation
\begin{equation}
 \frac{\pd^2 w}{\pd x^2}+\frac{\pd^2 w}{\pd y^2}=0.
\label{eq:laplace}
\end{equation}
The boundary conditions are equivalent to the transverse case, now applied to the velocity component $w$
\begin{align}
w&=N \delta(x)\frac{\pd w}{\pd y} &\for |x|<\frac{b}{2},\quad y&=0 \label{eq:bcreel1l}\\
w&=0 &\for |x|>\frac{b}{2},\quad y&=0.  \label{eq:bcreel2l}
\end{align}

We can seek the solution to the above problem in terms of an analytic function $W_\mr l(\mathfrak{z})$. The subscript l stands for the longitudinal flow direction. By definition, both the real and imaginary part of $W_\mr l(\mathfrak{z})$ fulfill the Laplace equation independently, hence we choose
\begin{equation}
w=\imag(W_\mr l(\mathfrak{z})).
\label{eq:wW}
\end{equation}
This leads to the boundary conditions
\begin{align}
\imag(W_\mr l)&=N \delta(x)\reel(W'_\mr l) &\for |x|&<\frac{b}{2},\quad y=0  \label{eq:bcimag1l}\\
\imag(W_\mr l)&=0 &\for |x|&>\frac{b}{2},\quad y=0  \label{eq:bcimag2l}
\end{align}
and the condition for the stress at infinity
\begin{equation}
\reel(W'_\mr l)= \frac{\tau_\infty}{\eta_1} \for y\rightarrow \infty.
\label{eq:bcimaginfl}
\end{equation}
Obviously, the structure of this boundary value problem is identical to the transverse case \eqref{eq:bcimag1} and \eqref{eq:bcimag2}. $W_\mr l(\mathfrak{z})$ directly relates to $W(\mathfrak{z})$, since both functions are analytic and obey boundary conditions differing only by constant factors. Goursat's theorem can therefore also be understood as a relation between the transverse and the longitudinal flow.

\begin{figure}
\centering
	\includegraphics{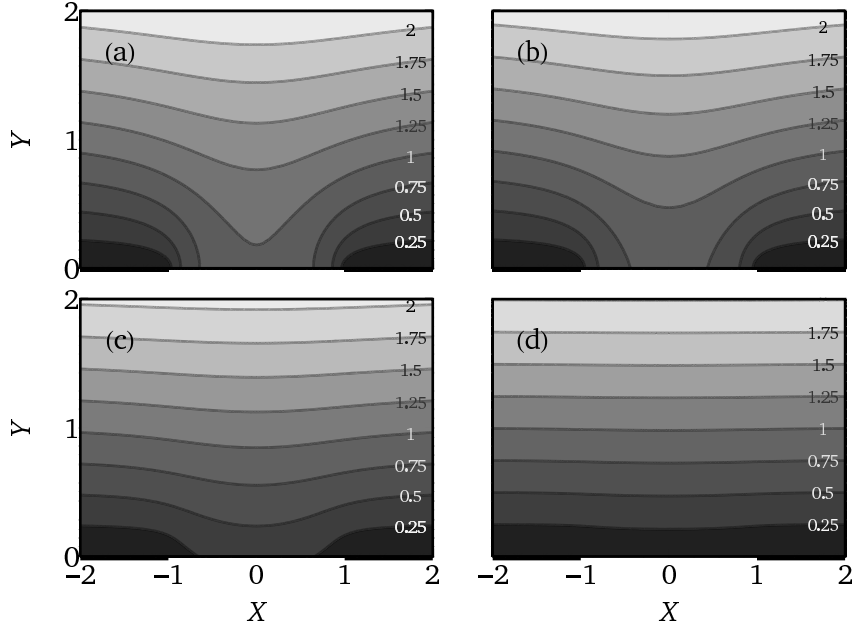}
	\caption{Longitudinal flow velocity $w/w_\mr{p,max}$ for $D=0.25$ and varying viscosity ratio: a) $N=100$, b) $N=10$, c) $N=1$, d) $N=0.1$}
	\label{fig:stroml_long}
\end{figure}

Thus, the solution process of the longitudinal case follows exactly the procedure outlined above. This calculation (for intermediate steps see appendix \ref{appA}) leads to a velocity
\begin{equation}
w=\frac{\tau_\infty}{\eta_1} \frac{1}{1+2 D N}\left(y+2 D N \imag{\sqrt{(x+\ii y)^2-\frac{b^2}{4}}}\right).
\end{equation}
The velocity isolines corresponding to several viscosity ratios and the same cavity geometry as in the previous section are shown in fig. \ref{fig:stroml_long}. Analogously, $w_\mr{p,max}$ is the maximum velocity at $y=0$ for perfect slip.
We observe the same flow enhancing effect for large $N$ as for the transverse case. Still, velocities are higher for the longitudinal flow, since $w_\mr{p,max}=2 u_\mr{p,max}$. This is in accordance with previous results \citep{philip1972}.

\section{Conclusions}
In this paper, we have investigated the flow over a rectangular cavity which contains a second immiscible fluid. Such a situation may arise in numerous practical applications, for example in situations in which a gas flows along along a rough surface with liquid-filled roughness features or when water flows along a superhydrophobic surface with entrapped gas bubbles.

In the calculation, the cavity is modeled as a boundary condition to the outer flow. The boundary condition appears in the form of a locally varying slip length, the functional form of which was determined by referring to previous results from lid-driven cavity flow. Up to now, analytical expressions for the flow field over such a cavity were known only for some limiting cases like a perfect, viscosity-free fluid filling the cavity or an infinitely deep cavity. We have achieved to formulate the problem in such a way that the flow over the cavity is directly related to the cavity geometry and the viscosity of the cavity fluid. The solution was obtained as an explicit analytical expression for the flow field and can be evaluated for arbitrary viscosities and arbitrary aspect rations of the cavity. We studied the flow in transverse and longitudinal direction. Both cases are connected via Goursat's theorem. The influence of the viscosity of the cavity medium on the outer flow was demonstrated, naturally being stronger the lower the viscosity. All results agree well with the known limiting cases. The analytical results have been compared to numerical calculations. The agreement between both is extremely good. \\

We kindly acknowledge support by the Cluster of Excellence 259 'Center of Smart Interfaces' and the 'Graduate School of Computational Engineering' at TU Darmstadt, both funded by the German Research Foundation (DFG).

\appendix
\section{}\label{appA}
The mixed boundary value problem
\begin{align*}
\imag(W_\mr l)&=N \delta(x)\reel(W'_\mr l) &\for & |x|<\frac{b}{2},\quad y=0 \\
\imag(W_\mr l)&=0 &\for & |x|>\frac{b}{2},\quad y=0 \\
\reel(W'_\mr l)&= \frac{\tau_\infty}{\eta_1} &\for & y \rightarrow \infty
\end{align*}
is solved by the method of \citet{sotkilava1974}.

The auxiliary differential equation
\begin{equation}
F_\mr l (\mathfrak{z})=W'_\mr l (\mathfrak{z})-\frac{1}{2 D N \sqrt{\mathfrak{z}^2-\frac{b^2}{4}}} W_\mr l (\mathfrak{z})
\label{eq:Fl}
\end{equation}
transforms the above problem to
\begin{align}
\reel(F_\mr l (\mathfrak{z}))&=0 &\for & |x|<\frac{b}{2},\quad y=0 \label{eq:bcf1l}\\
\imag(F_\mr l(\mathfrak{z}))&=0 &\for & |x|>\frac{b}{2},\quad y=0 \label{eq:bcf2l}\\
\reel(F_\mr l(\mathfrak{z}))&=\frac{\tau_\infty}{\eta_1}\left(1-\frac{1}{2 D N}\right)  &\for & y\rightarrow \infty&
\end{align}
The solution for $F_\mr l(\mathfrak{z})$ is given by the homogeneous part of the Keldysh-Sedov formula \eqref{eq:keldyshsedov}. Determining the coefficients according to the boundary conditions yields
\begin{equation}
F_\mr l (\mathfrak{z})=\frac{\tau_\infty}{\eta_1}(1-\frac{1}{2 D N})\frac{\mathfrak{z}}{\sqrt{\mathfrak{z}^2-\frac{b^2}{4}}}.
\end{equation}
Equation  \eqref{eq:Fl} can then be solved to give
\begin{equation}
W_\mr l(\mathfrak{z})=\frac{\tau_\infty}{\eta_1} \frac{ \mathfrak{z}+ 2 D N \sqrt{\mathfrak{z}^2-\frac{b^2}{4}}}{1+2 D N}.
\end{equation}

\bibliographystyle{jfm}

\bibliography{literatur_kavitaet}

\end{document}